# Novel Mechanism to Defend DDoS Attacks Caused by Spam

Dhinaharan Nagamalai
*Wireilla Net Solutions Inc, Chennai, India*

Cynthia Dhinakaran and Jae Kwang Lee
*Department of Computer Engineering, Hannam University, South Korea*

*Abstract*

*Corporate mail services are designed to perform better than public mail services. Fast mail delivery, large size file transfer as an attachments, high level spam and virus protection, commercial advertisement free environment are some of the advantages worth to mention. But these mail services are frequent target of hackers and spammers. Distributed Denial of service attacks are becoming more common and sophisticated. The researchers have proposed various solutions to the DDOS attacks. Can we stop these kinds of attacks with available technology? These days the DDoS attack through spam has increased and disturbed the mail services of various organizations. Spam penetrates through all the filters to establish DDoS attacks, which causes serious problems to users and the data. In this paper we propose a novel approach to defend DDoS attack caused by spam mails. This approach is a combination of fine tuning of source filters, content filters, strictly implementing mail policies, educating user, network monitoring and logical solutions to the ongoing attack. We have conducted several experiments in corporate mail services; the results show that this approach is highly effective to prevent DDoS attack caused by spam. The novel defense mechanism reduced 60% of the incoming spam traffic and repelled many DDoS attacks caused by spam.*

## 1. Introduction

Email is a source of communication for millions of people world wide [8]. But spam is abruptly disturbing the email users by eating their resource, time & money. In Internet community the spam has always been considered as bulk and unsolicited. Spam mails accounts for 80% of the entire mail traffic. Many researchers have proposed different solutions to stop the spam. But the effort has become a drop of water in the ocean. No matter how hard, spammers always find new ways to deliver spam mail to the user's inbox. Of late the spammers target the mail servers to disturb the activities of organizations which results in economic and reputation loss. The DDoS attack is a common mode of attack to cripple the particular server. The spammers take DDoS attack in their arms to disturb the mail servers. This paper is going to study the DDoS attacks through spam mails. We proposed a multi layer approach to defend the DDoS attack caused by spam mails. We implemented this methodology in our mail system and monitored the results. The result shows that our approach is very effective to defend DDoS attack caused by spam.

E-mail life cycle : The composed mail in the source machine will be handover to the Message Transfer Agent (MTA). The MTA will find the destination machine with the help of DNS server and relay the mail to the destination systems MTA [14]. The MTA at the destination machine delivers the mail to the destination user's mail box. The machines between source and destination will act as intermediate machines for the data transfer called relay. MTA relay mail between each other using the Simple Mail Transfer Protocol.

Most of the corporate mail services are pretty faster and sophisticated than other free mail services. The corporate mail services deliver mails faster and it provides a facility to attach bigger size files and unlimited storage facilities. To deliver mails faster, the server generally jumps most of the time consuming spam protection tests. To attach the big size files it has to bypass the content filter settings. This makes the corporate mail servers vulnerable to spam mail which ultimately causing DDOS attacks. We proposed a multi layer approach to defend the DDoS attack caused by spam mails. We implemented this methodology in our mail system and monitored the results. The result shows that our approach is very effective to defend DDoS attack caused by spam.

The rest of the paper is organized as follows. Section 2 discusses related work. Section 3 provides background on mechanism of DDOS attack through spam and the effects. In section 4, we describe our mechanism to defend the attack. Section 5 provides data Collection and experimental results. We conclude in section 6.

## 2. Related work

In [1] luiz, Crsitino presented an extensive study on characteristics of spam traffic in terms of email arrival process, size distribution, the distributions of popularity and temporal locality of email recipients etc., compared with legitimate mail traffic. Their study reveals major differences between spam and non spam mails. In [2] examines the use of DNS black lists. They have examined seven popular DNSBLs and found that 80% of the spam sources are listed in some DNSBL. [4] Presented a comprehensive study of clustering behavior of spammers and group based anti spam strategies. Their study exposed that the spammers has demonstrated clustering structures. They have proposed a group based anti spam frame work to block organized spammers. [5] Presented a network level behavior of spammers. They have analyzed spammers IP address ranges, modes & characteristics of botnet. Their study reveals that blacklists were remarkably ineffective at detecting spamming relays. Their study states that to trace senders the internet routing structure should be secured. [9] Presented a comprehensive study of spam and spammers technology. His study reveals that few work email accounts suffer from spam than private email. To the best of our knowledge our study is a first paper, comprehensively studying the DDoS attacks by spam.

## 3. Mechanism of DDoS attacks through spam

Distributed Denial of Service (DDoS) attack is a large scale, coordinated attack on the availability of services at a victim system or network resource [3]. DDOS attack through spam mail is one of the new versions of common DDoS attack. In this type, the attacker penetrates the network by a small program attached to the spam mail. After the execution of the attached file, the mail server resources will be eaten up by mass mails from other machines in the domain results denial of services. The working scenario of this attack is explained in fig 1. The attackers take maximum effort to pass through the spam filters and deliver the spam mail to the user's inbox. Here the hackers are doing enough to make the mail recipient to believe that the spam mail is from the legitimate user. The attackers use fake email ids from victim domains to penetrate the network. The spam mail had sent in the name of Network administrator/well wisher of the victim or boss of the organization. Note that the spam mail does not have the signature.

The spam contains small size of .exe file as an attachment (for example update.exe). The attackers used double file extension to confuse the filter (Update_KB2546_*86.BAK.exe (140k)) and user. The attachment size ranges from 140 to 180 KB. Mostly the spam mail asks the recipient to execute the .exe file to update anti virus software. Upon execution of the attachment, it will drop new files in windows folder and change the registry file, link to the attacker's website to download big programs to harm the network further. The infected machine collected email addresses through windows address book and automatically send mails to others in the same domain. Even if the users don't use mail service programs like Outlook express and others, it will send mails by using its own SMTP. Mostly this kind of spam mail attracts the group mail ids, and will send mails to groups. By sending mails to the group, it will spread the attack vigorously. If any of the users forward this mail to others it will worsen the situation. Ultimately the server will receive enormous request from others beyond its processing capacity. In this way it will spread the attack and results in a DDoS attack. After the first mail, for every minute it will send same kind of mail with different subject name & different contents to the group email ids. With in a day it will eat up server resources and end up in distributed denial of service attack. The names of the worms used in these kind of DDoS attacks are WORM_start.Bt,WORM_STRAT.BG,WORM_STRAT.BR, TROJ_PDROPPER.Q. Upon execution, these worms dropped files namely serv.exe, serv.dll, serv.s, serv.wax, E1.dll, rasaw32t.dll etc.

DDoS malware cause direct and indirect damage by flooding specific targets [14]. Mass mailers and network worms cause indirect damage when they clog mail servers and network bandwidth. In Network, It will consume the network bandwidth and resources, causing slow mail delivery further resulting Denial of service. The server will be down due to enormous request from clients and bulk mail processing.

## 4. Proposed Defense Mechanism

We proposed a multi layer approach to defend the DDoS attack caused by spam mails [Figure 2]. We implemented this approach in our mail system and monitored the results. The result shows that our approach is very effective. The approach has six layers as shown in Figure4. This approach is a combination of fine tuning of source filters, content filters, network monitoring policy, general email policies, educating the user & timely logical solutions of the network administrator. Fine tuning of source filters reject the incoming connections before the spam mail delivery. The content filters analyses the contents of the mails and blocks the incoming unwanted mails. Network monitoring approach provides general solution to identify the attacks prior to the attack and also during the attack. Business houses should educate the user about possible attack scenarios & reacting ways to it. The logical solutions of the network administrator play an important role during the attack period and even post attack period. The combination of these layers provides best methodology to stop the DDoS attacks established though spam mails.

### 4.1. Source Filters

Currently, There is a prediction that the spam will be 70% of the email traffic in 2007[1]. There are lot of source filters are available in real time. But by simply enabling all the filters will not help to prevent the attacks. It will slow down the mail delivery process. So the fine tuning of filter is an important to handle the attacks. The figure .3 shows the structure of the filters.

**Bayesian Filter:** Bayesian filtering is one of the effective filtering technologies used by most of the antispam software developers [9]. This filter works based on the mathematical theorem of Bayes a British mathematician. Anti spam developers have developed various algorithms by modifying the Bayes theorem to effectively filter the spam.

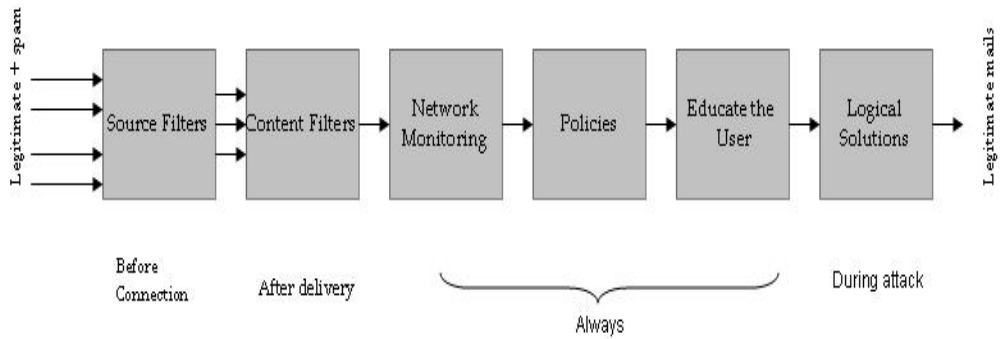

Figure 1. Defense Mechanism

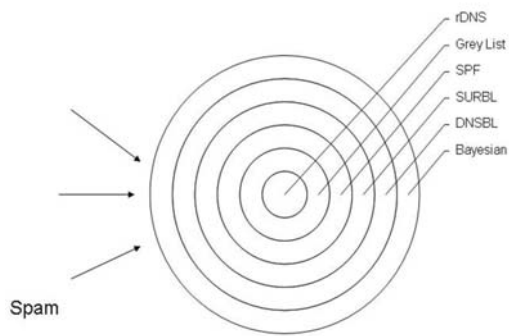

Figure 2. Combination of Source Filters

Anti spam companies have developed various algorithms by modifying the Bayes theorem to effectively filter the spam. In Bayes methodology, the system develops two tables from the contents of incoming spam mail & out bound legitimate mails. The tables referred as a dictionary. Each word from an incoming new mail will be compared to the spam mail table and legitimate mail table or dictionary. For incoming mail words, the probability value is calculated based on the number of occurrences of particular word in spam mail table & legitimate mail table.

For example the word "Viagra" occurs 400 times in 3000 spam mails and in 5 out of 300 legitimate mails the probability would be .889 [400/3000]/ [5/300+400/3000]. It will perform the same operation for all words in incoming mails. The mail is classified as spam, if the calculated probability is higher than a given threshold mostly above 0.5. The normal threshold value ranges from .7 and above for a corporate mail server. Based on the threat level the administrator can change the threshold value. After the classification of the mail as a spam, the contents of the spam will be added to the dictionary. It will be useful for the future calculation. In this way the system will learn the latest technologies used by the spammers. The administrator can enable or disable the learning from the spam and outbound mails. Most of the filters offer an administrator to select the number of words for dictionary. We recommend 50000 words are recommended for small & medium size mail server. If you increase the dictionary size, the lookup time or processing time will increase causing the delay in mail delivery. The system will take two weeks to build a spam word table but some of the filters use static tables. Bayesian filter is the most important and successful antispam method[9].The spammers learned how to pass this filter to deliver spam into inbox even it is powered with learning. Enhancements may be useful [10] but an ever increasing the quality of filter is impossible since the dictionary size & the content of the mail is limited. "Mail, server, report, firewall, virus, windows, customer, support" are some of the common tokens listed in legitimate mail table and also used by the attacker in a given sample mail. The server will send a regular report titled "server report" to the network administrator. So the probability of being a spam for this mail is very less results safe delivery to the user's inbox. After the attack the system will add these tokens to the spam tokens table. Meantime the spammers will come with new techniques. The legitimate mail word table is a standard one not enables for learning. The hackers can get the data and use it to bypass the filter. Bayesian filtering is effective but it can not filter 100% of the spam. With the combination of various filters Bayesian Filter will work more effective and overall performance of spam filtering will be increased.

**DNSBL:** Even though the spam generation is not accepted widely as a legal actively, 80% of the spam mail is generated by particular users. If we have the list of these spam generators IP addresses, we can effectively block the spam messages. DNSBL is based on the concept above said. DNS black hole list or black list is a well defined source filtering technology it works before delivering the mail to the user's inbox. The list publishes the list of IP addresses through DNS of massive spam generators. Lot of DNSBLs offers various list of IP addresses based on open relay, spam or virus source. The most widely used DNSBLs are spamhaus, spamcop,sorbs, abuseat,dsbl, rfc-ignorant etc., these DNSBLs list out thousands of IP addresses of spam generators. By blocking this well known IP addresses to deliver message we can effectively block the incoming spam traffic. There might be a overlapping of IP addresses in various Black lists. Legitimate Mail delivery will be delayed, if all the available DNS black lists included in the DNSBL settings to filter the spam. The IP address of the black listed IP addresses will change frequently based on their spam generating behaviors. Some DNSBLs will check the particular IP address regularly; if they stop the spamming activity, it will remove the particular IP address and add the new IP addresses of spammers [2].

If you enable "immediately reject the connections" from blacklisted server option, the connection will not be established to the particular spammer. Most of the mail servers provide the option to include more lists or delete the DNSBLs from the list. But the remaining 20% of the spam mail generators are not listed by any DNSBLs, still you have to depend on other filters or methodologies to stop the spam. Still there is no single Blacklist with all the complete spam generating IP addresses since the spammers will change their IP addresses frequently. Moreover these list providers are frequent target to hackers. The spammers used Mimail.E worm to perform Dos attack on spamhaus site. In 2003, Spamhaus servers came under distributed Denial of Service (DDoS) attacks by thousands of virus-infected computers throughout the Internet [20]. In 2006 also the spamhaus servers are out of service due to DDoS attacks [25]. It is clear that the angry spammers are trying to stop the services of DNSBLs. These attacks clearly show the use more than one DNSBLs in the List. Even if one

DNS black list is out of service the mail server can manage with other lists. In recent days the DNSBL lookups are increased tremendously of total DNS lookups compared to 5 years before [1]. Nearly 80% of the spam generated by relays that appear in one at least one of eight major blacklists [4]. Fine tuning of multiple black lists is more effective than simply using all lists. The DNSBLs is not effective when the spam is being sent from larger set of IP addresses [2].

**SURBL:** SURBL Searches for URLs in incoming mails. SURBL is a collection of spam supported websites, domains, web servers. If there is any URL or IP address in the message, the system will contact the SURBL list to check whether the URL is listed. If the URL is listed in SURBLs, it blocks the messages. The available SURBL lists are sc.surbl.org, ws.surbl.org, ob.surbl.org, ab.surbl.org. multi.surbl.org is a combination of all the lists. If the system uses other SURBLs with multi.surbl.org, it will take long time to process the mail. If use only multi.surbl.org for SURBL check, and if the service is not available, no checks will be performed. We recommend using other four surbls rather than multi.surbl.org. The administrator can edit the list whenever a high rate of false positive is present [17].

In the mentioned DDoS attack through spam [in section 2], the worm downloaded malicious code from the following websites.

http://www2.{BLOCKED}tinmdesachlion.com
http://www3.{BLOCKED}tinmdesachlion.com
http://www4.{BLOCKED}tinmdesachlion.com
http://www6.{BLOCKED}tinmdesachlion.com

If SURBL was enabled, there was less possibility of the attack. This kind of URL based filter is very effective against the DDoS attack since these references are faked websites. Some attacker includes multi URLs to confuse the filters. For multi domain messages, it is hard to determine the real spam domain among all the domains [10]. The combination of checking SURBL database with other filters is a best way to defend the DDoS attacks.

**Sender Policy Framework:** Sender Policy Framework reject message if SPF test is fail or soft fail [23]. Sender address forgery is a big threat to the users as well as the entire network. In the attack mentioned in the section 2, all the users received mails from the unknown person within their organization. The attacker's mail id is a fake, it has victims domain name. That is why most of the users obeyed the instruction and executed the file attachment leading to the DDoS attack. We can stop this kind of forgery by SPF (Sender Policy Framework).The current version of SPF — called SPFv1 or SPF Classic [13]. The Sender Policy Framework allows you to check whether a particular email sender is forged or not. Most of today's spammers use forged email addresses to hide their identity. SPF requires that the organization of the sender has published its mail server in an SPF record. If you receive a mail from a user, you can check that mail is coming from the particular organization by the sender's IP address. The SPF record will inform the receiver whether the user is allowed to use their network or not [15]. If the organization recognizes particular machine, it passes the test. Otherwise it is an attacker or a spammer. There are to kind of fails like fail and soft fail.

**Grey Listing:** Grey listing is a simple technique to fight against spam [18]. It will reject all incoming mails from unfamiliar IP addresses with an error code. The mail server records the combination of sender, recipient id & IP address. If the same sender is trying to send the mail after 10 seconds to 12 hours, the server will check for the combination in its record, if it matches, it will allow the sender to deliver the message. This is based on assumption that the spammers will not try again but legitimate users. But spammers learned this technology & how to bypass. But results show that there is substantial reduction of spam after the implementation of grey listing. The old version of Grey list used to accept the second mail after 4 hours [19]. But the legitimate user faces delay in mail delivery.

**Reverse DNS:** The incoming system should have rDNS ie. The sending system should give domain name and IP address to prove that is from the legitimate user. Most of the spam doesn't have reverse DNS [12]. Rejecting all incoming mails without rDNS is an effective way to filter the spam. "Reject message if sending server IP does not have a reverse DNS entry", "Reject message if the reverse DNS entry does not match Helo host" are two options supported by most mail services. SPF & rDNS are useful to filter the spam into some extends.

### 4.2. Content Filters

Once cleared from the SMTP server, the sender is allowed to deliver the message headers and body of the mail [12]. By carefully checking each and every word of the header and contents still we can block the spam. Most spam headers try to confuse the filters. Spammers will use recognizable words as a subject and clear from address. If the incoming mail has particular content or subject, the content filter will stop the mail delivery. Most of the spam caused DDOS attack has subjects like test, server report, status, helo etc; In this case the attacker carefully selected the words to avoid the content filtering. "Server report" is a word used by servers to send report to the administrator. The content filter blocks the mail which has some specific words like Viagra, ViAgRa, install updates, customer support service etc., Multiple words separated by comma, space are allowed in content filters to search the mail contents .

Most of the manuals say that the blocking of particular id & IP address is not useful [26]. But such activity highly reduces the spam mail delivery. Even the entire domain blocking is highly advisable to stop further spam delivery or attack. But in this DDoS attack, further attacks will be from its own domain mail ids except the first mail. Blocking own mail id is highly impossible. Another option of content filter is if you know your regular contacts, you can block all other mail senders. We can block all mails not send by particular user. If the administrator doesn't want to receive from other mail ids rather than own domain ids, he can block all the incoming and outgoing mails by enabling option "Block incoming message not sent by ". Content filters allow the admin to block all the mails with big size attachments. For example he can block all incoming mails with the attachment size is more than 100 KB. In this case the attacker can not deliver worms which are bigger in size. This will completely block the DDOS attack through this way. Moreover chances of DDoS attack caused by dumping of larger size files will be eliminated at greater extend. Denial of service will be delayed from starting time of the attack; meantime the admin can take other steps to defend the DDoS attack.

The content filters can block the mails with particular type of files as an attachment [12]. Most of these worms have .exe as an extension. If the filters block all the incoming mails with attachment of exe files, can stop the incoming DDOS attackers spam. But these days the spammers learnt this and started to send the attachments with double extension like update.doc.exe, update.txt.exe etc.  It will confuse the filter and will be delivered to mailbox. To avoid this, content filter is providing the options like *.*.exe. By this kind of technique the content filters are helpful to defend the DDoS attacks through spam. Before the attack, the content filters can not identify the contents, headers, subject, and the attachment size & file extension of the incoming mails. Anyway this spam will take some time to result the DDoS attack. After the first one or two mails the administrator can identify these item and block by using content filters. The experience of the administrators can play a wide role in this filter. While all of the filters are effective to some extends, a combination of these filters will effectively stop the DDoS attacks through spam mails. Even if it is pass through one filter, it will be blocked by another one. After the initial attack the combination of content filters play an important role to defend the DDoS attacks resulted by spammer..

### 4.3. Policies

Mail is the primary source of communication between all employees at an organization. Therefore it is appropriate that an email-etiquette be established to distinguish between what is Push vs. Pull information. As any organization of any size, it needs an agreed upon system of sending, sorting and utilizing files in their mail server. The type and number of emails / files sent via mail has increased exponentially over the past few years. If the server reaches its capacity levels that cause significant delay in email ultimately results DOS. The policy helps to avoid the DDoS attack kind of situations.

The users required to use signatures or set default signatures to all ids in the domain. So the users can easily differentiate spam and legitimate mail. Mostly the attackers' mail will not contain any signatures. Encourage user to use specific words in certain place. For example if the organization is religious organization, force all the users to use particular word to identify the mail is from legitimate user. For example if all the users are using "In His Service" instead of "regards" at the end of the mail. The attackers can not identify such words to make the user to believe the mail is from legitimate user.  Since most of the DDOS attacks through the spam mail takes an extra effort to make the user to believe the mails from legitimate user. Mostly the attack mails sent in the name of network administrator or head of the Institute. These unique code words will separate the spam from legitimate mails. The user will not open the mail or execute the attached file. It will help the network to fight against DDoS attacks through spam effectively.

### 4.4. Educate the user

The user's action during the attack or before the attack plays an important role to defend the DDoS attacks. So the users need to be educated how to behave generally and during the attack. The users have to be educated about spam mails and DDOS attacks. The users should be asked not to open or reply or forward or any kind of activities to the mails from unknown users. The user should inform the network administrator, if they responded to the spam in any method. The user can choose to flag spam so that the server knows to block it. The users should be asked not to use their work email addresses when registering in news groups and others. They should also be asked not to run any exe file or any file sent by email. Automatically deleting spam after particular day should be implemented. If not user should be advised to clean up their spam regularly. After the attack if spam mail exists with DDoS attack weapon, by error it can reappear and results DDoS attack. So the users should clear their old mail and spam regularly.

**4.5 Monitoring the Network**

To defend network against DDoS attack through spam requires real time monitoring of network wide traffic to obtain timely and important information. Monitoring the performance of network plays an important role to avert the DDOS attack [14]. Unusual activities can be detected, if the network is monitored by 24*7. If the speed of the mail service is down, we can assume that the server is processing a bulk data. Even the heavy regular network traffic causes the congestion; the administrator can regulate the data flow by his regular procedures to increase the speed. But during the attack, the net admin can not ease the data flow by his regular practices. It indicates that there is something wrong in the network. If the DDoS attack takes place automatically the mail server's speed will go down. Continuously monitoring the network performance is a useful practice to defend the attack. Two mail servers were monitored simultaneously with minimum of 200 mail users. In one domain the attacker launched DDoS attack through spam, because of continuous monitoring the net administrator marked the mail as a spam and deleted before spread to others. Absolutely the domain was escaped from the attack. In another domain for experiment, the spam was not marked and allowed to user's inbox. Some of the users responded to the spam by forwarding and executing the mail. Since this attack targeted only group ids, the mail server was out of service with in a day. Maintaining the history of network activities and network problems are useful to handle the current situation. Experiences are highly useful to design a solution to DDoS attacks through spam.

**4.6. Logical solutions**

Any attack can be handled with minimum impact by the network administrator's skills. After the attack, shutting down the server is not useful. The ways should be identified to change the path of the data dumping. The DDoS attack through spam mail targeted only group ids. So the mail service will become out of service very soon. But the wise net administrator can change all the group ids to new ids. For example allstaff@ABC.com can be changed in to all_staff@ABC.com. These group mail ids are converted into private users and not for public users. So the attacker is not allowed to send more mails. Since the incoming spam has diverted, all the spam mails stopped immediately. But already infected machines will give trouble to the particular users. The infected machines need to be removed from the network. In order to view the impact of the attack, these machines have to be analyzed. After the removal of worms from these machines, they can be allowed to join the network. There will be a logical solution to every attack, no need to be panic.

**5. Data Collection and Results**

We have conducted several experiments to measure the effectiveness of DNSBL, SURBL and the proposed defense mechanism. The test was conducted on client computers connected through local area network. The web server provides service to 200 users with 20 group email

IDs and 200 individual mail IDs. The speed of the Internet connection is 100 Mpbs for the LAN, with 20 Mbps upload and download speed (Due to security and privacy concerns we are not able to disclose the real domain name). Our dataset consists of the spam mails collected at a large spam trap. The trap is a collection of spam mails filtered by source, content filters, and other settings mentioned in this paper.

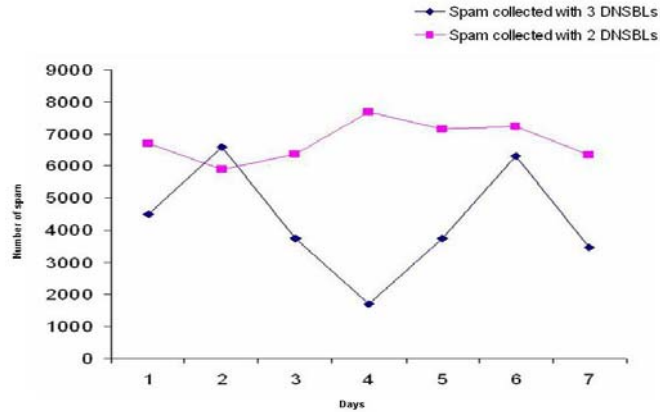

Figure3. Effectiveness of DNSBL

Several experiments were conducted to measure the effectiveness of DNSBL. Relays.ordb.org, bl.spamcop.net, sbl.spamhaus.org is a good combination to effectively filter the spam. For continues seven days we have included relays.ordb.org, bl.spamcop.net, sbl.spamhaus.org in DNS black lists and collected the spam. Then we have removed relays.ordb.org from DNSBL list and measured the spam trap. When we removed relays.ordb.org from the list, we observed 40~50% increase of spam traffic as shown in Figure3. Apart from this each user received 2~3 false negatives per day. The standard deviation of false negatives received by an end user for 7 days is shown in Figure4.

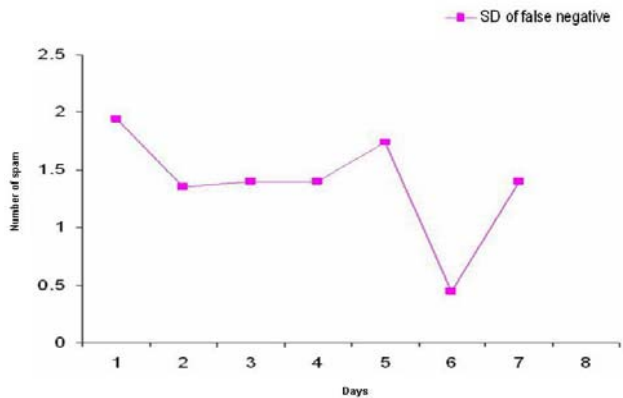

Figure4. Standard Deviation of false negative

We conducted several experiments to measure the effectiveness of SURBL. The test was conducted in the same network. To test the SURBL, we observed mail delivery for particular period of time (sessions). Each session is about 3 hours period of time. The experiment result shows that the effectiveness of the SURBL test. Our dataset consists of the spam mails collected at a large spam trap.

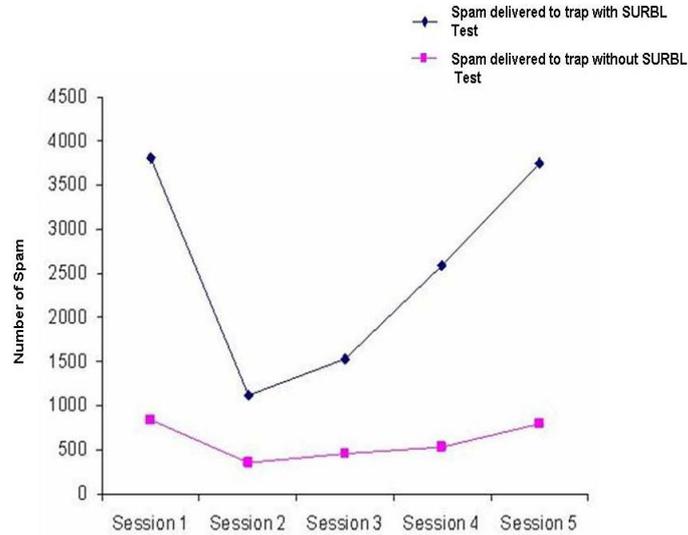

Figure5. SURBL test-Spam delivery

The number of spam had increased to the user's inbox when the SURBL test is not conducted to check the spam; at the same time the number spam has decreased to the spam trap. SURBL test was unchecked for five sessions. Most of the users received spam in their inbox during this test. The results are shown in the Figure5.

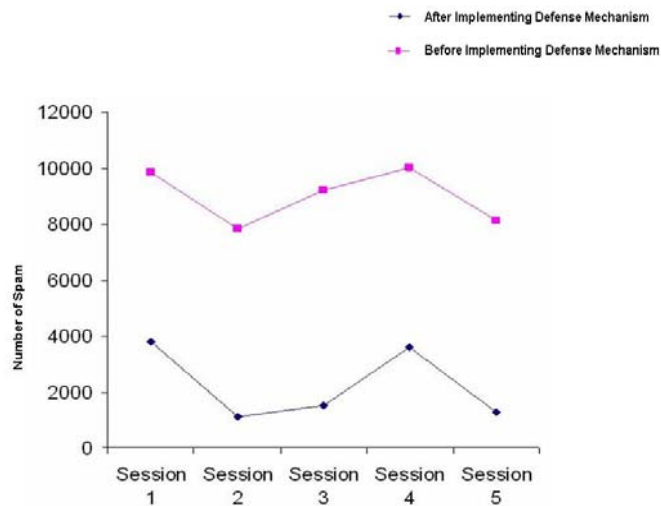

Figure6. Defense Mechanism effects

Several experiments were conducted to measure the effectiveness of the proposed defense mechanism. We observed the system for six months continuously. Our dataset consists of the spam mails collected at a large spam trap. The graph shows the number of spam received before and after implementing the defense mechanism. We have selected five sessions of data to display. As shown in Figure6, a session holds good for three hours. The graph shows that after implementing the defense mechanism the incoming spam has reduced by 50% to 60%. Our corporate mail service did not face any DDoS attack for past six months. We have observed that the individual users are not receiving more spam like before implementing the defense mechanism.

## 6. Conclusion

In this paper we have proposed a multi layer defense mechanism to defend the mail services from DDoS attacks caused by spam. Experimental results show that this system is highly effective and the mail service experiencing strong protection against DDoS attacks caused by spam. There is no single step solution to the DDoS attacks established through spam mails. Simply using various filters doesn't help to stop the possible DDoS attacks caused by spam. But fine tuning of filters mentioned in our mechanism prevented DDoS attacks through spam. The content filters clogged the attack by filtering the spam with unwanted contents and programs. Continuous monitoring of the network averted possible attacks and gave enough time to defend the attacks. Since the educated users are responding well to this kind of attacks, the attacks avoided in an efficient way. The policies prevented the spam mails by utilizing policies of using signatures, no bulk mails, and the limitations of usage of group ids. Last but certainly not the least, the logical solutions to these attacks plays an important role to stop the attacks.  The experiments show the effectiveness of SURBL to filter the spam. The experimental results show that there is 60% of reduction in spam traffic after implementing the defense mechanism. Also we didn't face DDoS attack through spam for past six months.

## 7. References

[1] Jae Yeon Jung, Emil sit: An empirical study of spam traffic and the use of DNS Black lists, ACM SIGCOMM Internet measurement conferences, pp 370-75, 2004.

[2] Anirudh Ramachandran, David Dagon, Nick Feamster: Can DNS-based Blacklists keep up with Bots, CEAS 2006, July 27-28, 2006.

[3] Guangsen Zhang, Manish Parashar: Cooperative mechanism against DDOS attacks, , SAM 2005, pg 86-96, June, 2005.

[4 Anirudh Ramachandran, Nick Feamster : Understanding the network level behaviour of spammers, SIGCOMM 06, September 11-16,2006.

[5] Ben Laurie, Richard Clayton: Proof of work proves not to work, WEIS04, Minneapolis MN, May 13-14, 2004.

[6] Ben Adida, David Chau, susan Hohenberger, Ronald L rivest: Lightweight Signatures for Email, DIMACS, 2006.

[7] David A Turner, Daniel M Havey: Controlling spam through Lightweight currency, ICCS-04, 2004.

[8] Carl Eklund: Spam -from nuisance to Internet Infestation, Peer to Peer and SPAM in the Internet Raimo Kantola's technical report, 126-134, 2004.

[9] Vladimir Mijatovic: Mechanism for detecting and prevention of email spamming, Peer to Peer and SPAM in the Internet Raimo Kantola's technical report, 135-145, 2004.

[10] Keno Albrecht, Nicolas Burri, Roger Wattenhofer: Spamato-An extendable spam filter system, Keno Albrecht, Nicolas Burri, Roger Wattenhofer, CEAS-05, July 2005.

[11] Jean-Marc Seigneur, Nathan Dimmock, Ciaran Bryce, Christain Damsgaard Jensen: Combating spam with TEA (Trustworthy email addresses) PST 2004, 47-58, 2004.

[12] Technical response to spam, Taughannok networks, Technical report, November 2003.

[13] Ralph F Wilson: SPF helps Legitimate E-mail get through spam filters, Web marketing today premium, Issue 85, November 2004.

[14] Jian Yuan, Kevin Mills: Monitoring the effect of Macroscopic Effect of DDOS flooding attacks, IEEE Transactions on Dependable and Secure Computing, Volume 2, No. 4, pp. 324-335, 2005.

[15] Sender Policy Framework, http://new.openspf.org/

[16] Spamhaus survives DDoS attack http://www.virus.org/news, September 2006.

[17] SURBL http://www.surbl.org/lists.html

[18] Grey Listing-http://greylisting.org/

[19] Greylisting FAQ https://hdc.tamu.edu/

[20] The Spamhous Project. www.spamhaus.org

[21] SpamAssassin ww.spamassassin.apache.org

[22] SpamCop http://spamcop.net

[23] Cloudmark, http://www.cloudmark.com

[24] Sender Policy Framework, http://new.openspf.org/

[25] Email Bombing and Spamming: www.cert.org/tech_tips/email_bombing_spamming.html

**INTENTIONAL BLANK**